\documentclass[numreferences]{kluwer}
\usepackage{amsfonts}
\usepackage{ams,amssymb,amsthm,comment}

\newtheorem{theorem}{Theorem}
\newtheorem{corollary}{Corollary}

\newcommand \eqref[1]{(\ref{#1})}

\begin{document}

\begin{opening}

\title{Stability of the (two-loop) Renormalization Group Flow for Nonlinear Sigma Models}

\author{Christine \surname{Guenther} \email{guenther@pacificu.edu}}
\institute{Department of Mathematics and Computer Science, Pacific University, Forest
Grove, Oregon, 97116, USA}

\author{Todd A. \surname{Oliynyk} \email{todd.oliynyk@sci.monash.edu.au}}
\institute{School of Mathematical Sciences, Monash University, VIC
3800, Australia}

\runningtitle{Stability of the two-loop Renormalization Group flow}
\runningauthor{Christine Guenther and Todd A. Oliynyk}

\begin{abstract}
We prove the stability of the torus, and with suitable rescaling, hyperbolic
space under the (two-loop) renormalization group flow for the nonlinear
sigma model. To prove stability we use similar techniques to \cite{GIK02},
where the stability of the torus under Ricci flow was first established. The
main technical tool is maximal regularity theory.
\end{abstract}


\classification{Mathematics Subject Classifications (2000)}{35K55, 53C80, 58Z05, 81T17}

\end{opening}

\section{Introduction}

The renormalization group flow equations for the world-sheet nonlinear sigma
models arise from quantizing the classical action
\[
S(x)=\frac{1}{4\pi \alpha ^{\prime }}\int_{\Sigma }\gamma ^{\alpha \beta
}g_{ij}(x)\partial _{\alpha }x^{i}\partial _{\beta }x^{j}dV(\gamma ),
\]%
where $\alpha ^{\prime }>0$ is the string coupling constant, $(\Sigma
,\gamma )$ is a $2$-dimensional Riemannian manifold (i.e. world sheet), $%
(M,g)$ is a $n$-dimensional Riemannian manifold (i.e. target space), and $x$
$:\Sigma $ $\rightarrow $ $M;$ $(\theta ^{1},\theta ^{2})$ $\mapsto $ $%
(x^{1}(\theta )$ $,\ldots $ $,x^{n}(\theta ))$ is a map. To (perturbatively)
quantize the classical theory, a momentum cutoff $\Lambda >0$ must be
introduced. This gives rise to a one parameter family of quantum field
theories indexed by the cutoff $\Lambda $. The target space metric becomes $%
\Lambda $ dependent and plays the role of the \textquotedblleft coupling
constants\textquotedblright . The requirement that the one parameter family
of field theories be equivalent on length scales $L\gg 1/\Lambda $ leads to
the Renormalization Group (RG) flow equations
\begin{equation}
\frac{\partial \;}{\partial \Lambda }g_{ij}=-\beta _{ij}^{g}\,.
\end{equation}%
In the perturbative regime ($\alpha ^{\prime }\ll 1$), the $\beta $%
-functions $\beta _{ij}^{g}$ can be expanded in powers of $\alpha ^{\prime }$
\cite{Frie85,JJM89}:
\begin{equation}
\beta _{ij}^{g}=\alpha ^{\prime }R_{ij}+\frac{{\alpha ^{\prime }}^{2}}{2}%
R_{iklm}R_{j}{}^{klm}+\mathrm{O}({\alpha ^{\prime }}^{3})\,.  \label{beta}
\end{equation}%
Here we are using $R_{ijk\ell }=g_{\ell m}R_{ijk}^{m}$ where $R_{ijk}^{\ell
}=\partial _{i}\Gamma _{jk}^{\ell }-\partial _{j}\Gamma _{ik}^{\ell }+\Gamma
_{jk}^{m}\Gamma _{im}^{\ell }-\Gamma _{ik}^{m}\Gamma _{jm}^{\ell }$.
Introducing \textquotedblleft time\textquotedblright\ by $t=-\ln (\Lambda )$%
, the RG flow equations become
\begin{equation}
\frac{\partial \;}{\partial t}g_{ij}=-\alpha ^{\prime }R_{ij}-\frac{{\alpha
^{\prime }}^{2}}{2}R_{iklm}R_{j}{}^{klm}+\mathrm{O}({\alpha ^{\prime }}%
^{3})\,.  \label{RG}
\end{equation}%
Truncating at the first order in $\alpha ^{\prime }$ gives
\begin{equation}
\frac{\partial \;}{\partial t}g_{ij}=-\alpha ^{\prime }R_{ij}.  \label{RG1}
\end{equation}%
In the perturbative regime, the first order truncation should provide an
acceptable approximation to the full RG flow. However, this is difficult to
quantify as a rigorous definition of the $\beta $-functions requires a
non-perturbative quantization of the nonlinear sigma model, and this has not
yet been shown to exist. As has been noted previously \cite{Lott86}, the
first order truncation (\ref{RG}) is \emph{Ricci flow}. It is easy to see
(e.g. by considering the sphere) that there exist many solutions to Ricci
flow which become singular in finite time. In \cite{Ham82} Hamilton showed
that at a finite singular time $T<\infty $, the Riemannian curvature blows
up, i.e. $\lim_{t\nearrow T}|Rm|_{g}=\infty $. Therefore for times near a
singular time for Ricci flow, from the perturbative expansion for the
beta-functions (\ref{beta}), it appears that the second order correction
term would dominate even for $\alpha ^{\prime }\ll 1$. This would signal a
regime change where the first order truncation (i.e. Ricci flow) is no
longer a valid approximation for the full RG flow. This suggests the
strategy of using the second order (two-loop) truncation
\begin{equation}
\frac{\partial \;}{\partial t}g_{ij}=-\alpha ^{\prime }R_{ij}-\frac{{\alpha
^{\prime }}^{2}}{2}R_{iklm}R_{j}{}^{klm}\,  \label{RG2}
\end{equation}%
as a measure of the error in approximating the full RG flow (\ref{RG}) by
Ricci flow (\ref{RG1}). A related motivation for studying (\ref{RG2}), as
discussed in \cite{Lott86}, is that in certain situations, it is enough to
consider the second order RG equations (\ref{RG2}) to establish the
existence of a continuum limit \cite{GK85}. Finally, we note that it is
tempting to view the higher order truncations of the RG flow equations (\ref%
{RG}) as natural modifications of Ricci flow. In this light, the $\beta $%
-function expansion (\ref{beta}) in powers of $\alpha ^{\prime }$ generates
specific modifications to Ricci flow. As with other equations that arise
from physical considerations, these equations may have \textquotedblleft
nice\textquotedblright\ properties, which we think, at the very least, is
worth investigating. However, we note that whereas (\ref{RG2}) can
be studied using parabolic techniques (at least for $\alpha ^{\prime }$
small enough),
the third and higher order correction would involve
polynomials of the curvature and its derivatives\cite{JJM89}. This would
mean that in trying to study the third and higher order equations as
evolution equations, some form of the Nash-Moser implicit function theorem
would be necessary to overcome the resulting loss of derivatives.

As a first step in the analysis of the second order RG flow (\ref{RG2}), we
prove the stability of the flat torus, and with suitable rescaling
hyperbolic space, under the flow. See sections \ref{torus} and \ref%
{hyperbolic} for the theorems and proofs. The main theorems are Theorem \ref%
{Torus} and Theorem \ref{modified hyperbolic}.

\section{Stability of the second order RG flow at $M=\mathbb{T}^{n}.$ \label%
{torus}}

Let $\bar{g}$ be a flat Riemannian metric on a closed $n-$dimensional
manifold $M$. In this section we show that there exists a neighborhood of $%
\bar{g}$ in an appropriate function space such that if $g_{0}$ is in this
neighborhood, then the solution $g(t)$ of the renormalization group flow
with $g(0)=g_{0}$ converges exponentially quickly to a centermanifold at $%
\bar{g}$ consisting entirely of flat metrics$.$ To do so we will apply
maximal regularity theory, as introduced by \cite{DG79}, and applied to
fully nonlinear equations in \cite{DL88}. See \cite{RFV2} for an
introduction to these methods in a geometric setting. The argument is
similar to that used to show the stability of the Ricci flow at flat metrics
\cite{GIK02}, and at hyperbolic metrics \cite{KY06}\footnote{%
See \cite{Ye93} for results on the stability of constant nonzero curvature
spaces under Ricci flow, using other means.}; however, the second order
renormalization group flow is no longer quasilinear, and so results for
fully nonlinear equations must be used (see the cross-curvature flow result
in \cite{KY06} for another fully nonlinear example). We will work with the
nested little H\"{o}lder spaces of 2-tensors, denoted $h^{i+\alpha }$, which
are the completion of the $C^{\infty }$ 2-tensors in the H\"{o}lder norm $%
\left\vert \left\vert \cdot \right\vert \right\vert _{i+\alpha }$. Letting $%
E_{0}=h^{0+\sigma }$, $E_{1}=h^{2+\sigma }$, $X_{0}=h^{0+\rho }$, and $%
X_{1}=h^{2+\rho }$, where $0<\sigma <\rho <1,$ one has
\[
X_{1}\subset E_{1}\subset X_{0}\subset E_{0},
\]%
with $X_{0}$ and $X_{1}$ the continuous interpolation spaces required to
apply maximal regularity theory (for an introduction to these spaces, their
connection with the more familiar real interpolation spaces, and a
justification of the above inclusion, see the beginning of Section 3.3 \cite%
{GIK02}). We prove the following theorem:

\begin{theorem}[Stability of n-dimensional Torus]
\label{Torus}Let $\bar{g}$ be a flat Riemannian metric on a closed $n-$%
dimensional manifold $M$, $n\geq 3,$ and let $g(t)$ satisfy the second order
RG flow
\begin{eqnarray}
\frac{\partial }{\partial t}g &=&-2Rc(g)-\frac{\alpha ^{\prime }}{2}Rm^{2}
\label{2nd order RF} \\
g(0) &=&g_{0}.  \nonumber
\end{eqnarray}%
There exists a ball $B_{r}(\bar{g})$ $\subset h^{2+\rho }$, $r>0$, such that
if $g_{0}\in B_{r}(\bar{g}),$ then the solution $g(t)$ of (\ref{2nd order RF}%
) converges exponentially quickly to a flat metric.
\end{theorem}

\begin{pf}
Following \cite{DT03}, in order to obtain a strictly parabolic equation, for
any positive definite $2$-tensors $g,u$ we define a vector field
\begin{equation}
W_{u,g}^{i}=-g^{ij}u_{jk}^{-1}g^{kl}g^{pq}\left( \nabla _{p}u_{ql}-\frac{1}{2%
}\nabla _{l}u_{pq}\right) ,  \label{W_u,g}
\end{equation}%
and a modified flow
\begin{equation}
\frac{\partial }{\partial t}g=-2Rc(g)+L_{W_{u,g}}g-\frac{\alpha ^{\prime }}{2%
}Rm^{2}(g),  \label{DeTurck-RGF}
\end{equation}%
where $L_{W_{u,g}}g$ is the Lie derivative of $g$ in the direction $W_{u,g}.$
We shall call this the second order DeTurck-RG flow. If $g$ is a solution of
(\ref{DeTurck-RGF}), then $\phi _{t}^{\ast }g$ is a solution of (\ref{2nd
order RF}), where $\phi _{t}$ is the family of diffeomorphisms generated by
integrating the vector field $-W_{u,g}$. When $u=\bar{g}\ $we see that $\bar{%
g}$ is also a fixed point of (\ref{DeTurck-RGF}). By the standard first
variation formula $\frac{d}{d\varepsilon }Rc(g+\varepsilon h)|_{\varepsilon
=0}$ (see Theorem 1.174 \cite{B87}), the linearization of the first
two terms of the right hand side at $\bar{g}$ is given by
\begin{equation}
D(-2Rc(g)+L_{W_{\bar{g},g}}g)|_{g=\bar{g}}h=\bar{\Delta}_{L}h,  \label{Lin}
\end{equation}%
where
\begin{equation}
\bar{\Delta}_{L}h_{ij}=:\bar{\Delta}h_{ij}+2\bar{R}_{kijl}h^{_{^{{kl}%
}}}-g^{kl}\bar{R}_{il}h_{kj}-g^{kl}\bar{R}_{jl}h_{ik}  \label{Lich}
\end{equation}%
is the Lichnerowicz Laplacian with respect to the metric $\bar{g},$ and $%
\bar{\Delta}$ denotes the usual Laplacian with respect to $\bar{g}$ (see
Chapter 3.3 in \cite{CK04} for complete details of the calculation) . We
denote $\bar{R}_{ijkl}=Rm(\bar{g})_{ijkl}$ and $\bar{R}_{ij}=Rc(\bar{g})_{ij}
$. So letting $u=\bar{g}$, the linearization of (\ref{DeTurck-RGF}) at a
flat metric $\bar{g}$ is
\begin{equation}
\frac{\partial }{\partial t}h=\bar{\Delta}h=:A_{\bar{g}}h.
\label{Linearization}
\end{equation}

The existence of an exponentially attractive centermanifold is a consequence
of Theorem 3.3 in \cite{DL88}, and so we verify its hypotheses. It is
convenient to rewrite equation (\ref{DeTurck-RGF}) as
\[
\frac{\partial }{\partial t}g=A_{\bar{g}}g(t)+G(g(t)),
\]%
where $A_{\bar{g}}\ $is the linearization of the right hand side of (\ref%
{DeTurck-RGF}). The hypotheses of the theorem are:\newline
\begin{tabular}{l}
$A_{\bar{g}}:X_{1}\rightarrow X_{0}$ is sectorial (and extends to sectorial $A_{\bar{g}%
}:E_{1}\rightarrow E_{0}$), \\
$G\in C^{1}(O,X_{0})$, where $O\subset X_{1}$ is a neighborhood of $\bar{g}$,
\\
$G(\bar{g})=0$, $\;\;G^{\prime }(\bar{g})=0$.%
\end{tabular}

We recall that an operator being sectorial roughly means that the spectrum is bounded in a
wedge in a left half plane, and its resolvent is uniformly bounded on the
comlementary right half plane. By definition, the space $h^{2+\alpha }$
consists of 2-tensors whose second derivatives are in $h^{0+\alpha }$; as
was shown in Lemma 3.4 \cite{GIK02}, by standard Schauder estimates the
operator $A_{\bar{g}}$ $:X_{1}\rightarrow X_{0}$ is sectorial; continuous Frechet
differentiability of $G$ and the existence of $O$ can be checked as in
Section 4 of \cite{KY06}. Since
\[
\langle A_{\bar{g}}h,h\rangle =\int \langle \bar{\Delta}h,h\rangle
=-\left\vert \left\vert \bar{\nabla}h\right\vert \right\vert ^{2},
\]%
the kernel of $A_{\bar{g}}$ consists of all parallel 2-tensors, and is an $%
\frac{n(n+1)}{2}$ dimensional subspace of the tangent space of symmetric
2-tensors.Therefore the hypotheses of the theorem are satisfied, and there
is an $r>0$ such that if $g_{0}\in B_{r}(\bar{g})\subset h^{2+\rho }$, there
exists a centermanifold at $\bar{g}$ of dimension at most $\frac{n(n+1)}{2}$
that is exponentially attractive \ for times $t\leq \tau (r)$. Flat metrics
are fixed points of the flow, and so since they are exponentially attracted
to the centermanifold, they must lie on the centermanifold. Since the space
of flat metrics on the torus is $\frac{n(n+1)}{2}$ dimensional, the
centermanifold must consist precisely of flat metrics.

We can in fact obtain a global result, using Corollary 9.1.6 in \cite{Lun95}
to obtain long term existence. Consider a solution $g$ of (\ref{DeTurck-RGF}%
). By definition (\ref{W_u,g}), if $g$ converges exponentially to a flat
metric, then $W_{\bar{g},g}$ converges exponentially to zero. The curvature
terms also converge exponentially to zero, and so the right hand side of (%
\ref{DeTurck-RGF}) converges to zero exponentially and $\left\vert \frac{%
\partial }{\partial t}g\right\vert \leq Ce^{-\omega t}$, where $\omega >0.$
(See also the proof of Theorem 3.7 in \cite{GIK02}.) We therefore have
global existence and convergence to a flat metric.

We have shown exponential convergence of solutions of DeTurck-RG flow, but
we have exponential convergence of the second order RG flow as well. The
idea is that if the norm of a vector field\ $W$ decays exponentially, then
the diffeomorphisms $\phi _{t}$ generated by the vector field converge
exponentially to a fixed diffeomorphism. (To see this just calculate the
length of the integral curves. See also Lemma 3.5 and Proposition 3.6 in
\cite{GIK02}.) Since the solutions of (\ref{DeTurck-RGF}) and (\ref{2nd
order RF}) are related by the pullback $\phi _{t}^{\ast }$, the theorem is
proved.
\end{pf}

\section{\protect\bigskip Stability of the modified second order RG flow at $%
M=$ $\mathbb{H}^{n}$ \label{hyperbolic}}

We next consider the stability of a closed hyperbolic manifold. Let $\bar{g}$
be a Riemannian metric of constant negative curvature $K\ $on a closed $n-$%
dimensional manifold $M$ with $n\geq 3$. In this case $\bar{g}$ is not a
fixed point of the second order RG flow (\ref{2nd order RF}), and so we will
instead consider an equation that is related to the renormalization group
flow only by diffeomorphism and rescaling. To proceed, let $g$ satisfy the
original RG flow
\[
\frac{\partial }{\partial t}g=-2Rc(g)-\frac{\alpha ^{\prime }}{2}Rm^{2}(g)
\]
for $0<t<T$. For a constant $c>0$, let $\sigma (t)=c(t+1)$ and $\tau (t)=%
\frac{1}{c}\log (t+1)$, so that $\frac{\partial }{\partial t}\tau =\frac{1}{%
c(t+1)}$. Define $\hat{g}(\tau )$ by
\[
g(t)=\sigma (t)\phi _{t}^{\ast }\hat{g}(\tau (t)),
\]
where $W$ is the vector field (\ref{W_u,g}), and $\frac{\partial }{\partial t%
}\phi _{t}=-W_{u,\hat{g}}$. Then
\begin{eqnarray*}
\frac{\partial }{\partial t}g &=&\frac{\partial }{\partial t}(\sigma (t)\phi
_{t}^{\ast }\hat{g}(\tau (t)) \\
&=&c\phi _{t}^{\ast }(\hat{g}(\tau (t))+c(t+1)\left[ \phi _{t}^{\ast }(\frac{%
\partial \hat{g}}{\partial \tau }\frac{\partial \tau }{\partial t})+\frac{%
\partial }{\partial s}|_{s=0}(\phi _{t+s}^{\ast }(\hat{g}(\tau (t))\right] \\
&=&\phi _{t}^{\ast }\left( c\hat{g}(\tau (t))+\frac{\partial \hat{g}}{%
\partial \tau }-L_{W_{u,\hat{g}}}\hat{g}(\tau (t))\right) .
\end{eqnarray*}%
Since
\begin{eqnarray*}
\frac{\partial }{\partial t}g &=&-2Rc(g)-\frac{\alpha ^{\prime }}{2}Rm^{2}(g)
\\
&=&\phi _{t}^{\ast }\left( -2Rc(\hat{g})-\frac{\alpha ^{\prime }}{2ce^{c\tau
}}Rm^{2}\left( \hat{g}\right) \right) ,
\end{eqnarray*}%
$\hat{g}$ satisfies the evolution equation%
\begin{equation}  \label{HRGF}
\frac{\partial }{\partial \tau }\hat{g}=-2Rc(\hat{g})+L_{W_{u,\hat{g}}}\hat{g%
}-\frac{\alpha ^{\prime }}{2ce^{c\tau }}Rm^{2}\left( \hat{g}\right) -c\hat{g}%
,
\end{equation}%
for $0\leq \tau <\frac{1}{c}\log (T+1)$. This equation has a time-dependent
coefficient $\frac{\alpha ^{\prime }}{2ce^{c\tau }}$, and so is no longer
autonomous; however, since the coefficient decays exponentially we can
handle this by considering instead the system
\begin{equation}  \label{ERGF}
\frac{\partial }{\partial \tau }\left(
\begin{array}{c}
\hat{g}_{ij} \\
v%
\end{array}%
\right) =\left(
\begin{array}{c}
-2\hat{R}_{ij}+L_{W_{u,\hat{g}}}\hat{g}_{ij}-v^{2}\hat{R}_{iklm}\hat{R}%
_{j}{}^{klm}-c\hat{g} \\
-\frac{c}{2}v%
\end{array}%
\right) .
\end{equation}
Notice that we can explicity integrate the $v$ equation to get $v=v_{0}e^{-%
\frac{c}{2}\tau }$. This shows, in particular, that if $(\hat{g}%
(\tau),v(\tau))$ is a solution to (\ref{ERGF}), then $g(\tau)$ will solve (%
\ref{HRGF}) with $\alpha^{\prime }$ defined via the formula
\[
\alpha^{\prime} = 2c v_0^2.
\]

Since $\hat{g}=\bar{g}$ is a metric of constant curvature $K<0$, we observe
that
\begin{eqnarray*}
\bar{R}_{ijkl} &=&K(\bar{g}_{il}\bar{g}_{jk}-\bar{g}_{ik}\bar{g}_{jl}), \\
\bar{R}m_{ij}^{2} &=&2K^{2}(n-1)\bar{g}_{ij}, \\
\bar{R}_{ij} &=&K(n-1)\bar{g}_{ij}.
\end{eqnarray*}%
The Lichnerowicz Laplacian
is
\begin{equation}
\bar{\Delta} _{L}h_{ij}=\bar{\Delta} h_{ij}+2KH\bar{g}_{ij}-2nKh_{ij},  \label{K-Lich}
\end{equation}
where $H=\bar{g}^{ij} h_{ij}$.
Setting
\[
c=-2K(n-1),
\]%
we then get that $(\bar{g},0)$ is a fixed point for (\ref{ERGF}). By (\ref%
{Lin}) and (\ref{K-Lich}), it is clear that the linearization of (\ref{ERGF}%
) at $(\hat{g},v)=(\bar{g},0)$ is given by
\[
\partial _{t}\left(
\begin{array}{c}
h \\
w%
\end{array}%
\right) =\left(
\begin{array}{cc}
\bar{\Delta} _{L}-c & 0 \\
0 & -\frac{c}{2}%
\end{array}%
\right) \left(
\begin{array}{c}
h \\
w%
\end{array}%
\right) .
\]%
Letting $A_{(\bar{g},0)}(h,w)$ denote the righthand side of the
linearization, we see that
\begin{eqnarray*}
\langle A_{(\bar{g},0)}(h,w),(h,w)\rangle  &=&\int (h,w)\left(
\begin{array}{cc}
\bar{\Delta} _{L}-c & 0 \\
0 & -\frac{c}{2}%
\end{array}%
\right) \left(
\begin{array}{c}
h \\
w%
\end{array}%
\right)  \\
&=&\int (\bar{\Delta _{L}}-c)h_{ij}h^{ij}-\frac{c}{2}w^{2} \\
&=&\int (\bar{\Delta} h_{ij}+2KHg_{ij}-2nKh_{ij}-ch_{ij})h^{ij}-\frac{c}{2}w^{2} \\
&=&-\left\vert \left\vert \bar{\nabla} h\right\vert \right\vert ^{2}+2K\left\vert
\left\vert H\right\vert \right\vert ^{2}-(2nK+c)\left\vert \left\vert
h\right\vert \right\vert ^{2}-\frac{c}{2}\left\vert \left\vert w\right\vert
\right\vert ^{2}.
\end{eqnarray*}%
Using the Koiso Bochner formula \cite{K79}
\begin{equation}
\left\vert \left\vert \bar{\nabla}h\right\vert \right\vert ^{2}=\frac{1}{2}%
\left\vert \left\vert \bar{T}\right\vert \right\vert ^{2}+\left\vert
\left\vert \bar{\delta}h\right\vert \right\vert ^{2}-nK\left\vert \left\vert
h\right\vert \right\vert ^{2}+K\left\vert \left\vert H\right\vert
\right\vert ^{2}
\end{equation}%
where $\bar{T}_{ijk} = \bar{\nabla}_k h_{ij} -\bar{\nabla}_i h_{jk}$,
yields
\begin{eqnarray*}
\langle A_{\bar{g},0}(h,w),(h,w)\rangle  &=&-\frac{1}{2}\left\vert
\left\vert \bar{T}\right\vert \right\vert ^{2}-\left\vert \left\vert \bar{\delta}
h\right\vert \right\vert ^{2}+K(n-2)\left\vert \left\vert h\right\vert
\right\vert ^{2}+K\left\vert \left\vert H\right\vert \right\vert ^{2} \\
&&+2K(n-1)\left\vert \left\vert w\right\vert \right\vert ^{2}.
\end{eqnarray*}%
This shows that the spectrum of $A_{(\hat{g},0)}$ is
strictly negative for $n\geq 3$. Since the Lichnerowicz Laplacian is
self-adjoint and elliptic, the resolvent bound follows again from standard
Schauder estimates (see e.g. Lemma 3.4 in \cite{GIK02}).

Thus, by Theorem 9.1.7 in \cite{Lun95}, one has exponential convergence of (%
\ref{ERGF}) for initial data in an appropriate neighborhood of $(\bar{g},0)$%
. Since we have an explicit formula for $v$, letting $\alpha ^{\prime
}=2cv_{0}^{2}$ we have proved the following theorem:

\begin{theorem}[Stability of modified RG flow for n-dimensional hyperbolic
space]
\label{modified hyperbolic}Let $\bar{g}$ be a Riemannian metric of constant
curvature $K<0$ on a closed $n-$dimensional manifold $M$, $n\geq 3$, and let
$\hat{g}(\tau )$ satisfy the modified DeTurck second order RG flow%
\begin{eqnarray}
\frac{\partial }{\partial \tau }\hat{g} &=&-2Rc(\hat{g})+L_{W_{\bar{g},\hat{g%
}}}\hat{g}(\tau )-\frac{\alpha ^{\prime }}{2ce^{c\tau }}Rm^{2}\left( \hat{g}%
\right) -c\hat{g}(\tau ),  \label{ModScalDiffeo} \\
\hat{g}(0) &=&\hat{g}_{0},  \nonumber
\end{eqnarray}%
where $c=-2K(n-1)$. Then for sufficiently small $\alpha ^{\prime }$, there
exists a ball $B_{r}(\bar{g})$ $\subset h^{2+\rho }$ such that if $g_{0}\in
B_{r}(\bar{g}),$ then the solution $\hat{g}(\tau )$ of (\ref{ModScalDiffeo})
exists for all time and converges exponentially quickly to a metric of
constant negative curvature.
\end{theorem}

We have seen that $\hat{g}(\tau )$ converges to one of constant negative
curvature, but what happens to the original solution of the second order RG
flow? Recall that
\[
g(t)=\sigma (t)\phi _{t}^{\ast }\hat{g}(\tau (t))
\]
is a solution of the second order RG flow. The scaling factor $\sigma $
blows up, but the vector field $W_{\bar{g},g}$ converges exponentially to
zero. Therefore, one has the following corollary:

\begin{corollary}
\label{thm hyperbolic}Let $\bar{g}$ be a Riemannian metric of constant
curvature $K\,<0$ on a closed $n-$dimensional manifold $M$, $n\geq 3$, and
let $g(t)$ satisfy the second order RG flow
\begin{eqnarray}
\frac{\partial }{\partial t}g &=&-2Rc(g)-\frac{\alpha ^{\prime }}{2}%
Rm^{2}\left( g\right)  \label{RGF b} \\
g(0) &=&g_{0},  \nonumber
\end{eqnarray}%
Then for sufficiently small $\alpha ^{\prime }$, there exists a ball $B_{r}(%
\bar{g})$ $\subset h^{2+\rho }$, $r\,>0$, such that if $g_{0}\in B_{r}(\bar{g%
})$, then the solution $g(t)$ of (\ref{RGF b}) exists for all time and
becomes more homogeneous in time.
\end{corollary}

\section{Discussion \label{discussion}}

We have established the stability of the torus, and with suitable rescaling
hyperbolic space, under the second order (two-loop) RG flow (\ref{RG2}).
This shows that in a neighborhood of a flat, or with suitable rescaling a
hyperbolic metric, the qualitative behavior of the first and second order RG
flow equations are the same. In particular, this suggests that in these
situations Ricci flow is a good approximation to the full RG flow. One
immediate application of this result is to produce approximate solutions to
the RG flow that involve order $\alpha ^{\prime 2}$ corrections, and for
which the Perelman type entropy proposed by Tseytlin \cite{T07} (see also
\cite{OSW05,OSW06,OSW07}) is monotone, at least to order $\alpha ^{\prime 2}$%
.

The results contained in this article are only a first step in the analysis
of the second order RG flow (\ref{RG2}). Clearly, it would be of interest to
identify initial data which generate solutions for which the first and
second order equations differ significantly. This could signal a breakdown
of the perturbative regime and therefore would be of physical interest. It
could also be of interest mathematically as the most likely mechanism for
the breakdown would be the development of large curvature along the flow.
Depending on the behavior of the solutions to the second order RG flow, this
may (or may not) lead to new applications in differential geometry.

\begin{acknowledgements}
This work began while both authors were at the Albert-Einstein-Institute (AEI).
We thank the AEI for its hospitality, and the director Gerhard Huisken of the Geometric Analysis and
Gravitaton group for creating such a supportive and stimulating work
environment.
\end{acknowledgements}

\bigskip

\end{document}